\documentclass[12pt]{article}

\usepackage{amssymb}
\makeatletter \@addtoreset{equation}{section} \makeatother

\def\ftoday{{\sl {Le \number\day \space\ifcase\month
\or janvier\or f\'evrier\or mars\or avril\or mai \or juin\or
juillet\or ao\^ut\or septembre\or octobre \or novembre \or
d\'ecembre\fi\space \number\year}}}
\def\ptoday{{\sl {\number\day \space de\space \ifcase\month
\or janeiro\or fevereiro\or mar{\c c}o\or abril\or maio \or
junho\or julho\or agosto\or setembro\or outubro \or novembro \or
dezembro\fi\space de\space \number\year}}}
\def\gtoday{{\sl {Den \number\day. \ifcase\month
\or Januar\or Februar\or M\"arz\or April\or Mai \or Juni\or
Juli\or August\or September\or Oktober \or November \or
Dezember\fi\space \number\year}}}
\def\today{{\sl {\ifcase\month
\or January\or February\or March\or April\or May \or June\or
July\or August\or September\or October \or November \or
December\fi \space\number\day,\space
                                            \number\year}}}

\newcommand{\XI}{\XI}

\newcommand{\sla}{\raise.15ex\hbox{$/$}\kern -.57em}
\newcommand{\Sla}{\raise.15ex\hbox{$/$}\kern -.70em}

\newcommand{\complex}{{\kern .1em {\raise .47ex
\hbox {$\scriptscriptstyle |$}}
    \kern -.4em {\rm C}}}
\newcommand{\real}{{{\rm I} \kern -.19em {\rm R}}}
\newcommand{\rational}{{\kern .1em {\raise .47ex
\hbox{$\scripscriptstyle |$}}
    \kern -.35em {\rm Q}}}
\renewcommand{\natural}{{\vrule height 1.6ex width
.05em depth 0ex \kern -.35em {\rm N}}}

\newcommand{\twiddle}{\lower.9ex\rlap{$\kern -.1em\scriptstyle\sim$}}

\newcommand{\eq}{\begin{equation}}
\newcommand{\eqn}[1]{\label{#1}\end{equation}}
\newcommand{\eea}{\end{eqnarray}}
\newcommand{\eqa}{\begin{eqnarray}}
\newcommand{\eqan}[1]{\label{#1}\end{eqnarray}}
\newcommand{\ba}{\begin{array}}
\newcommand{\ea}{\end{array}}
\newcommand{\eqac}{\begin{equation}\begin{array}{rcl}}
\newcommand{\eqacn}[1]{\end{array}\label{#1}\end{equation}}

\begin{document}

{\hfill\large {\bf break.tex}\ \ \ptoday}
{\hfill\parbox{45mm}{{hep-th/xxxxxxx\\

CBPF-NF-xxx/yy }} \vspace{3mm}

\begin{center}
{\Large \bf Off-Shell Extended Supersymmetries and
Lorentz-Violating Abelian Gauge Models}
\end{center}
\vspace{3mm}

\begin{center}{\large
Wander G. Ney$^{a,b,c,*}$, J. A. Helayel-Neto$^{b,c,**}$ and
Wesley Spalenza$^{b,c,d,***,}$\footnote{Supported by the Conselho
Nacional de Desenvolvimento Cient\'{\i}fico e Tecnol\'{o}gico CNPq
- Brazil.} } \vspace{1mm}

\noindent 
 $^{a}$Centro Federal de Educa\c{c}\~{a}o Tecnol\'{o}gica de Campos (CEFET) RJ-Brazil\\
$^{b}$Centro Brasileiro de Pesquisas F\'{i}sicas (CBPF) RJ-Brazil\\
$^{c}$Grupo de F\'{i}sica Te\'{o}rica Jos\'{e} Leite Lopes (GFT)\\
$^{d}$Scuola Internazionale Superiore di Studi Avanzati (SISSA)
Trieste-Italy.
\end{center}
\vspace{1mm}

{\tt E-mails: *wander@cbpf.br, **helayel@cbpf.br,
***spalenza@sissa.it}


\begin{abstract}
In this work, we propose the $N=2$ and $N=4$ supersymmetric
extensions realized off-shell of the Abelian gauge model with
Chern-Simons Lorentz-breaking term. We start with the theory in
$6$ and $10$ dimensions and reduce \`{a} la Scherk the space-like
coordinates to carry out the $D=5$ model in both cases. Then, we
reduce the fifth space-like coordinate using the Legendre
transformation technique for dimensional reduction. The last
reduction method provides us with auxiliary fields that yield the
superalgebra closed off-shell. Since the reduced bosonic
Lagrangians from $6D$ and $10D$ are the same as the $N=2$ and
$N=4$ SUSY-versions of the theory, respectively, we use the
superspace-superfield formalism in $N=1$ to achieve the
supersymmetric version of model.
\end{abstract}

\section{Introduction}

The field-theoretic models adopted to describe the fundamental
interactions among truly elementary particles have the Lorentz and
gauge symmetries as their main cornerstones. However, mechanisms
of breaking these symmetries have been proposed and discussed in
view of some phenomenological and experimental evidences
\cite{1,1b,1c,1d,1e}. Astrophysical observations indicate that
Lorentz symmetry may be slightly violated in order to account for
anisotropies. Then, one may consider a gauge theory where Lorentz
symmetry breaking may be realized by means of a term in the
action. A Chern-Simons-type term may be considered that exhibits a
constant background four-vector which maintains the gauge
invariance but breaks down the Lorentz space-time symmetry
\cite{1}.

In the context of supersymmetry (SUSY), a number of works
introduce the idea of Lorentz breaking in connection with SUSY: In
ref. \cite{3a}, SUSY was presented through the modification of the
algebra; in ref. \cite{3,3c}, one achieves the $N=1-$SUSY version
of the Chern-Simons term through a superspace-superfield
formalism; in ref. \cite{3b}, the authors adopt the idea of
Lorentz breaking operators. The $N=2$ and $N=4$ extended
supersymmetric version of the Lorentz breaking term have been
presented in \cite{3ca} using the ordinary dimension reduction
(\`{a} la Scherk) of the bosonic sector in 10 and 6 space-time
dimensions respectively and then using the superspace-superfield
formalism in $N=1$. In this last one, the extended
supersymmetrizations are realized on shell, i.e., the superalgebra
can not close without imposition of the equations of motion in 4D.
In the present paper, we propose the off-shell $N=2$ and $N=4$
supersymmetric version of the Abelian gauge sector added the
Chern-Simons Lorentz breaking term. The main idea is based on the
work of ref. \cite{2c}, where it was presented the $N=2$ and $N=4$
supersymmetric version realized off-shell for the Yang-Mills
theory. It was obtained through the dimensional reduction of the
$N=1$ supersymmetric model in components in 6 and 10 dimensions
respectively. In order to obtain the off-shell realization,
different of the refs. \cite {2b,3ca} presented on-shell, it
proceeds making the ordinary dimensional reduction of space-like
coordinates from the $N=1$ model in $6D$ and $10D$ to $5D$ and
than reducing the other space-like coordinate to $4D$ using the
Legendre transformation technique. The model receive some
auxiliary fields, that have close relations with the conjugate
momenta, that make superalgebra closed off-shell and the
translation in respect to the reduced coordinate (\`{a} la
Legendre) can be identified as central charge transformation.

In the ordinary dimension reduction, presented by Scherk, one
assumes that no field depends on the coordinate to be reduced,
then all derivatives with respect to this coordinates are taken to
be zero. The technique of reduction \`{a} la Legendre is not so
direct: its main idea is based on the Hamiltonian, not with
respect to the time, but with respect to the coordinate to be
reduced. In this technique, the fields and the Lagrangian are
dependent of the extra coordinate, but the reduced action does not
have this dependence. In this way, the fields have to obey the
equation of motion in the higher dimension that turns into
constraints of the reduced Lagrangian. In the first section, we
shall present these techniques and then we shall apply to a $D=6$
and a $D=10$ Abelian gauge model with Lorentz-violating
Chern-Simons term. In both applications, we reduce the space-like
coordinates \`{a} la Scherk to achieve the $D=5$ version of the
model, and then, we reduce more one space-like coordinate using
the Legendre transformation technique. In the second and the third
sections, we use the fact that the reduced Lagragians starting
with $D=6$ and $D=10$ model, obtained in the first section, are
the bosonic sectors of the off-shell $N=2$ and $N=4$
supersymmetric version of the model proposed, respectively. Once
the bosonic sector is identified, we adopt an $N=1$ -superfield
formalism to write down the gauge and the background
supermultiplets and then we set up
their coupling in terms of an $N=2$ and $N=4$ action realized in $N=1$%
-superspace. The result is projected out in component fields and
we discuss
the role of the background partners for the central charge of the $N=2$ and $%
N=4$ Lorentz broken action.

\section{The Scherk and Legendre procedures for dimensional reduction}

There are in the literature several techniques for dimensional
reduction, such as \`{a} la Scherk \cite{13,14}, \`{a} la Legendre
\cite{2c}, \`{a} la Kaluza-Klein \cite{15,16}, \`{a} la
Witten-Manton \cite{17,18,19}, and others. In this work, we shall
contemplate two of them: the technique \`{a} la Scherk and \`{a}
la Legendre to make the dimensional reduction of the Abelian gauge
model with a Lorentz violating term. This shall be useful to
achieve, in the next sections, the off-shell $N=2$ and $N=4$
supersymmetric version of this model. This model in $4D$ was
proposed by \cite{1}, and it is written as follows:
\begin{equation}
{\cal L}_{4}=-\frac{1}{4}F_{\mu \nu }F^{\mu \nu }+\varepsilon
^{\mu \nu \kappa \lambda }A_{\mu }\partial _{\nu }A_{\kappa
}t_{\lambda },  \label{0}
\end{equation}
where $t_{\lambda }$ is a constant vector that exist in the
background space-time, determining a preference direction, and
then, breaking the Lorentz symmetry.

In the first part of this section, we shall present the
dimensional reduction techniques \`{a} la Scherk (ordinary
technique) and \`{a} la Legendre (through Legendre
transformations). Then, in the second part, we shall apply these
techniques to make the dimensional reduction from $D=6$ to $D=4$
and from $D=10$ to $D=4$ for the $D=6$ and $D=10$ versions of
(\ref{0}). This procedure will provide the bosonic sector for the
$N=2-$ and $N=4-$SUSY generalization of this model, respectively.
Then, in the next sections, we will carry out these complete
off-shell supersymmetric model using the superfields formalism in
a $N=1$ superspace.

The ordinary reduction (\`{a} la Scherk) is very direct, it is
considered
that the fields do not depend on the extra coordinate to be reduced ($%
\partial _{4}A_{\hat{\mu}}=0$). In the reductions $D=6$ to $D=4$ and $D=10$
to $D=4,$ we will proceed reducing \`{a} la Scherk the space-like
coordinates to $D=5,$ and then, we will reduce the space-like coordinate $%
x^{4}$using the Legendre transformation technique.

In the Legendre transformation technique, it is supposed that the Lagrangian $%
{\cal L}_{4}$ has $x^{4}$-dependence ($\partial _{4}{\cal
L}_{4}\neq 0$), but the action $S_{4}$ does not depend on $x^{4}$
($\partial _{4}S_{4}=0).$ This kind of reduction is interesting
also because we suppose that physically the fields really depend
on the extra dimension.

After we obtain the Lagrangian in $5D,$ we make the dimensional
reduction with respect to a coordinate $x^{4}$ through the
Legendre transformation technique. It is based on the idea of the
Hamiltonian formalism. The Hamiltonian density is the Legendre
transformation of the Lagrangian (density) with respect to the
time and we know that the Hamiltonian density has time dependence
($\frac{d}{dt}{\cal H}\neq 0$), but the Hamiltonian is invariant
under time translations ($\frac{d}{dt}H=0$). The idea is to obtain
the Lagrangian ${\cal L}_{4}$ as a kind of ``Hamiltonian density''
with respect to the coordinate $x^{4}$. The Lagrangian ${\cal
L}_{4}$, unlike the \`{a} la Scherk method, depends on the
coordinates $x^{4}$:
\begin{equation}
\frac{\partial {\cal L}_{4}}{\partial x^{4}}\neq 0.
\end{equation}
However, the action must be independent of $x^{4}$:
\begin{equation}
\frac{\partial S_{4}}{\partial x^{4}}=0.
\end{equation}
In order to find the Lagrangian ${\cal L}_{4}$ from a Lagrangian ${\cal L}%
_{5}$ we make a procedure similar to obtain the Hamiltonian, but
unlike the
Hamiltonian, it is not with respect to the time, but to the coordinate $x^{4}$%
. Suppose the Lagrangian in 5 dimensions ${\cal L}_{5}$ doesn't
have explicit $x^{4}$ dependence (only implicity):
\begin{equation}
\frac{\partial {\cal L}_{5}}{\partial x^{4}}=\frac{\partial {\cal L}_{5}}{%
\partial A_{\hat{\nu}}}\frac{\partial A_{\hat{\nu}}}{\partial x^{4}}+\frac{%
\partial {\cal L}_{5}}{\partial \partial _{\hat{\mu}}A_{\hat{\nu}}}\frac{%
\partial \partial _{\hat{\mu}}A_{\hat{\nu}}}{\partial x^{4}},
\end{equation}
where $\hat{\mu}=0,1,2,3,4$. If we make the integration by parts
obtaining
\begin{equation}
\frac{\partial {\cal L}_{5}}{\partial x^{4}}=\left( \frac{\partial {\cal L}%
_{5}}{\partial A_{\hat{\nu}}}-\partial _{\hat{\mu}}\frac{\partial {\cal L}%
_{5}}{\partial \partial _{\hat{\mu}}A_{\hat{\nu}}}\right) \frac{\partial A_{%
\hat{\nu}}}{\partial x^{4}}+\partial _{\hat{\mu}}\left( \frac{\partial {\cal %
L}_{5}}{\partial \partial _{\hat{\mu}}A_{\hat{\nu}}}\frac{\partial A_{\hat{%
\nu}}}{\partial x^{4}}\right) .  \label{0a}
\end{equation}
Considering that the field equation in $5D$ is satisfied, then
\begin{equation}
\frac{\partial {\cal L}_{5}}{\partial A_{\hat{\nu}}}-\partial _{\hat{\mu}}%
\frac{\partial {\cal L}_{5}}{\partial \partial
_{\hat{\mu}}A_{\hat{\nu}}}=0 \label{2a}
\end{equation}
and (\ref{0a}) becomes
\begin{equation}
\frac{\partial {\cal L}_{5}}{\partial x^{4}}=\partial
_{\hat{\mu}}\left(
\frac{\partial {\cal L}_{5}}{\partial \partial _{\hat{\mu}}A_{\hat{\nu}}}%
\frac{\partial A_{\hat{\nu}}}{\partial x^{4}}\right) .
\end{equation}
Considering $\mu =0,1,2,4$, we have that
\begin{equation}
\frac{\partial {\cal L}_{5}}{\partial x^{4}}=\partial _{\mu }\left( \frac{%
\partial {\cal L}_{5}}{\partial \partial _{\mu }A_{\hat{\nu}}}\frac{\partial
A_{\hat{\nu}}}{\partial x^{4}}\right) +\partial _{4}\left(
\frac{\partial
{\cal L}_{5}}{\partial \partial _{4}A_{\hat{\nu}}}\frac{\partial A_{\hat{\nu}%
}}{\partial x^{4}}\right) .
\end{equation}
Then
\begin{equation}
\frac{\partial }{\partial x^{4}}\left( {\cal L}_{5}-\frac{\partial {\cal L}%
_{5}}{\partial \partial _{4}A_{\hat{\nu}}}\frac{\partial A_{\hat{\nu}}}{%
\partial x^{4}}\right) =\partial _{\mu }\left( \frac{\partial {\cal L}_{5}}{%
\partial \partial _{\mu }A_{\hat{\nu}}}\frac{\partial A_{\hat{\nu}}}{%
\partial x^{4}}\right) .
\end{equation}
If we integrate this expression in the 4 space-time volume:
\begin{equation}
\frac{\partial }{\partial x^{4}}\int d^{4}x\left( {\cal L}_{5}-\frac{%
\partial {\cal L}_{5}}{\partial \partial _{4}A_{\hat{\nu}}}\frac{\partial A_{%
\hat{\nu}}}{\partial x^{4}}\right) =\int d^{4}x\partial _{\mu }\left( \frac{%
\partial {\cal L}_{5}}{\partial \partial _{\mu }A_{\hat{\nu}}}\frac{\partial
A_{\hat{\nu}}}{\partial x^{4}}\right) ,  \label{2b}
\end{equation}
we notice that the term in the RHS of equation (\ref{2b}) must
turn into a hipersurface integral that must be zero, because
$A_{\hat{\nu}}$ must be fix. Then, we can define a Lagrangian
${\cal L}_{4}$ that can be thought of as the negative of the
``Hamiltonian'' with respect to the $x^{4}$ as:
\begin{equation}
{\cal L}_{4}=-{\cal H}_{5}={\cal L}_{5}-\frac{\partial {\cal L}_{5}}{%
\partial \partial _{4}A_{\hat{\nu}}}\frac{\partial A_{\hat{\nu}}}{\partial
x^{4}}.  \label{3}
\end{equation}
In this reduced Lagrangian, new auxiliary fields appear; they are
closely related to the canonical momenta
\begin{equation}
\pi ^{\hat{\nu}}=\frac{\partial {\cal L}_{4}}{\partial \partial _{4}A_{\hat{%
\nu}}}.
\end{equation}
These auxiliary fields have the role to close the off-shell
superalgebra in one dimension less.

We notice that to define (\ref{3}) as a Lagrangian in $4D,$ the
equations of motion in $5D$ (\ref{2a}) must be satisfied. It
becomes a constraint in $5D$ that determine how the physical and
auxiliary fields transform under $x^{4}$ translations. This
$x^{4}$ translations correspond to the central-charge
transformations in the reduced superalgebra.

\subsection{The chain of reductions $6D\rightarrow 5D\rightsquigarrow
4D\,\,\,\,$\protect\footnote{%
The symbol $\rightarrow $ means that the dimensional reduction is
performed through the technique \`{a} la Scherk, while the symbol
$\rightsquigarrow $ stands for the reduction \`{a} la Legendre.}}

In \cite{2c}, it was shown the attainment of the off-shell
$N=2-$SUSY generalization for the Yang-Mills model applying the
two techniques of
dimensional reduction mentioned for the $N=1-$SUSY model in components in $%
D=6$. Here, we shall apply these dimensional reduction techniques
for a $D=6$ version of the bosonic model (\ref{0}). The reduced
Lagrangian is the bosonic sector of the off-shell $N=2-$SUSY
generalization of this model. As proposed in \cite{3ca}, the $D=6$
version of the Lagrangian (\ref{0}) can be given by:
\begin{equation}
{\cal L}_{6}=-\frac{1}{4}F_{\dot{\mu}\dot{\nu}}F^{\dot{\mu}\dot{\nu}}+\frac{1%
}{3!}\varepsilon ^{\dot{\mu}\dot{\nu}\dot{\kappa}\dot{\rho}\dot{\lambda}\dot{%
\sigma}}A_{\dot{\mu}}\partial _{\dot{\nu}}A_{\dot{\kappa}}T_{\dot{\lambda}%
\dot{\rho}\dot{\sigma}},  \label{1}
\end{equation}
where $\dot{\mu}=0,1,2,3,4,5.$

The dimensional reduction of a space-like coordinate $x^{5}$ \`{a}
la Scherk is done considering no dependence of the fields in
respect to this coordinates. This reduction takes the form as
follows:
\begin{equation}
{\cal L}_{5}=-\frac{1}{4}F_{\hat{\mu}\hat{\nu}}F^{\hat{\mu}\hat{\nu}}+\frac{1%
}{2}\partial _{\hat{\mu}}\varphi \partial ^{\hat{\mu}}\varphi +\frac{1}{2}%
\varepsilon ^{\hat{\mu}\hat{\nu}\hat{\kappa}\hat{\lambda}\hat{\rho}}A_{\hat{%
\mu}}\partial _{\hat{\nu}}A_{\hat{\kappa}}R_{\hat{\lambda}\hat{\rho}}+\frac{1%
}{3!}\varepsilon ^{\hat{\mu}\hat{\nu}\hat{\kappa}\hat{\lambda}\hat{\rho}}A_{%
\hat{\mu}}\partial _{\hat{\nu}}\varphi T_{\hat{\kappa}\hat{\lambda}\hat{\rho}%
},  \label{2}
\end{equation}
where $\hat{\mu}=0,1,2,3,4$,\thinspace \thinspace \thinspace
\thinspace
\thinspace \thinspace $\varepsilon ^{\hat{\mu}\hat{\nu}\hat{\kappa}\hat{%
\lambda}\hat{\rho}5}\equiv \varepsilon ^{\hat{\mu}\hat{\nu}\hat{\kappa}\hat{%
\lambda}\hat{\rho}}$ and we redefined the gauge fields as
\[
(A_{\dot{\mu}})\rightarrow (A_{\hat{\mu}},\,\varphi ),
\]
and the background fields as
\[
(T_{\dot{\lambda}\dot{\rho}\dot{\sigma}})\rightarrow (R_{\hat{\lambda}\hat{%
\rho}},\,T_{\hat{\kappa}\hat{\lambda}\hat{\rho}}).
\]
We can notice that in $5D$, the theory has one vector and one
scalar gauge fields; one rank-2 and one rank-3 (that could be
considered its dual that is rank-2) background fields. The next
step is to make the dimensional
reduction $5D\rightsquigarrow 4D$ \`{a} la Legendre of the $5D$ Lagrangian (%
\ref{2}). Substituting (\ref{2}) in
\begin{equation}
{\cal L}_{4}={\cal L}_{5}-\frac{\partial {\cal L}_{5}}{\partial
\partial
_{4}A_{\hat{\nu}}}\frac{\partial A_{\hat{\nu}}}{\partial x^{4}}-\frac{%
\partial {\cal L}_{5}}{\partial \partial _{4}\varphi }\frac{\partial \varphi
}{\partial x^{4}},
\end{equation}
and expliciting the index $4,$ we obtain the Lagrangian
\begin{eqnarray}
{\cal L}_{4} &=&-\frac{1}{4}F_{\mu \nu }F^{\mu \nu
}+\frac{1}{2}F_{4\mu }F^{4\mu }+\partial _{\mu }\phi F^{4\mu
}+\frac{1}{2}\partial _{\mu }\varphi
\partial ^{\mu }\varphi -\frac{1}{2}\partial _{4}\varphi \partial ^{4}\varphi
\label{4} \\
&&+\varepsilon ^{\mu \nu \kappa \lambda }A_{\mu }\partial _{\nu
}A_{\kappa }p_{\lambda }+\varepsilon ^{\mu \nu \kappa \lambda
}A_{\mu }\partial _{\nu }\phi R_{\kappa \lambda
}+\frac{1}{2}\varepsilon ^{\mu \nu \kappa \lambda }A_{\mu
}\partial _{\nu }\varphi S_{\kappa \lambda }+\phi \partial _{\mu
}\varphi t^{\mu },  \nonumber
\end{eqnarray}
where $\mu =0,1,2,3$, $\varepsilon ^{\mu \nu \kappa \lambda
4}\equiv \varepsilon ^{\mu \nu \kappa \lambda },$ and we redefine
the gauge fields as
\[
(A_{\hat{\mu}},\,\varphi )\rightsquigarrow (A_{\mu },\,\phi
,\,\varphi ),
\]
and the background fields as
\begin{equation}
(R_{\hat{\lambda}\hat{\rho}},\,T_{\hat{\kappa}\hat{\lambda}\hat{\rho}%
})\rightsquigarrow (p_{\lambda },\,R_{\lambda \rho },\,S_{\kappa
\lambda },\,t^{\mu }).
\end{equation}
Now, we can notice that the theory in $4D$ has a vector and two
scalars gauge fields; two vectors (a vector was given by the dual
of one rank-3 tensor) and two rank-2 tensor background fields.

The canonical reduction with respect to $x^{4}$ is:
\begin{eqnarray}
\pi (A_{\mu }) &=&\frac{\partial {\cal L}_{5}}{\partial \partial _{4}A_{\mu }%
}=-F^{4\mu }+\frac{1}{2}\varepsilon ^{\mu \nu \kappa \lambda
}A_{\nu
}R_{\kappa \lambda }, \\
\pi (\varphi ) &=&\frac{\partial {\cal L}_{5}}{\partial \partial
_{4}\varphi }=\partial ^{4}\varphi -A_{\mu }t^{\mu }.
\end{eqnarray}
We define the auxiliary fields as
\begin{equation}
F_{4\mu }\equiv \eta _{\mu },\,\,\,\,\,\,\,\,F_{4}^{\,\,\,\,\mu
}=-F^{4\mu }\equiv \eta ^{\mu },\,\,\,\,\,\,\,\,\partial
_{4}\varphi =-\partial ^{4}\varphi \equiv G.
\end{equation}
This will be necessary to close our future superalgebra off-shell.
Replacing these definitions to the Lagrangian (\ref{4}), one gets:
\begin{eqnarray}
{\cal L}_{4} &=&-\frac{1}{4}F_{\mu \nu }F^{\mu \nu
}-\frac{1}{2}\eta _{\mu }\eta ^{\mu }-\eta ^{\mu }\partial _{\mu
}\phi +\frac{1}{2}\partial _{\mu
}\varphi \partial ^{\mu }\varphi +\frac{1}{2}GG  \label{5} \\
&&+\varepsilon ^{\mu \nu \kappa \lambda }A_{\mu }\partial _{\nu
}A_{\kappa }p_{\lambda }+\varepsilon ^{\mu \nu \kappa \lambda
}A_{\mu }\partial _{\nu }\phi R_{\kappa \lambda
}+\frac{1}{2}\varepsilon ^{\mu \nu \kappa \lambda }A_{\mu
}\partial _{\nu }\varphi S_{\kappa \lambda }+\phi \partial _{\mu
}\varphi t^{\mu }.  \nonumber
\end{eqnarray}
The $4D$ Lagrangian (\ref{5}) will be invariant off-shell only if
the constraints imposed by the equations of motion in $5D,$
relative to the
Lagrangian (\ref{2}), were satisfied. The equation of motion for $A_{\hat{\nu%
}}$ is presented as follows:
\begin{equation}
\partial _{\hat{\mu}}F^{\hat{\mu}\hat{\nu}}+\frac{1}{2}\varepsilon ^{\hat{\nu%
}\hat{\mu}\hat{\kappa}\hat{\lambda}\hat{\rho}}F_{\hat{\mu}\hat{\kappa}}R_{%
\hat{\lambda}\hat{\rho}}+\frac{1}{3!}\varepsilon ^{\hat{\nu}\hat{\mu}\hat{%
\kappa}\hat{\lambda}\hat{\rho}}\partial _{\hat{\mu}}\varphi T_{\hat{\kappa}%
\hat{\lambda}\hat{\rho}}=0.  \label{6}
\end{equation}
Considering $\hat{\nu}=4$, we have
\begin{equation}
\partial _{\mu }\eta ^{\mu }-\frac{1}{2}\varepsilon ^{\mu \nu \kappa \lambda
}F_{\mu \nu }R_{\kappa \lambda }-\partial _{\mu }\varphi t^{\mu
}=0, \label{7}
\end{equation}
and for $\hat{\nu}=\nu ,$%
\begin{equation}
\partial _{4}\eta ^{\nu }=\partial _{\mu }F^{\mu \nu }+\frac{1}{2}%
\varepsilon ^{\nu \mu \kappa \lambda }F_{\mu \kappa }p_{\lambda }+\frac{1}{2}%
\varepsilon ^{\nu \mu \kappa \lambda }\partial _{\mu }\varphi
S_{\kappa \lambda }-\varepsilon ^{\nu \mu \kappa \lambda }\eta
_{\mu }R_{\kappa \lambda }+Gt^{\mu }.  \label{8}
\end{equation}
The equation of motion for $\varphi $ is presented as below
\begin{equation}
\partial _{\hat{\mu}}\partial ^{\hat{\mu}}\varphi -\frac{1}{3!}\varepsilon ^{%
\hat{\mu}\hat{\nu}\hat{\kappa}\hat{\lambda}\hat{\rho}}\partial _{\hat{\mu}%
}A_{\hat{\nu}}T_{\hat{\kappa}\hat{\lambda}\hat{\rho}}=0.
\end{equation}
Taking the component $\hat{\mu}=4$, this equation looks as
\begin{equation}
\partial _{4}G=\partial _{\mu }\partial ^{\mu }\varphi -\frac{1}{4}%
\varepsilon ^{\mu \kappa \lambda \rho }F_{\mu \nu }S_{\lambda \rho
}-\eta _{\mu }t^{\mu }.  \label{9}
\end{equation}

Notice that the constraint (\ref{7}) can simplify the Lagrangian
to
\begin{eqnarray}
{\cal L}_{4} &=&-\frac{1}{4}F_{\mu \nu }F^{\mu \nu
}-\frac{1}{2}\eta _{\mu
}\eta ^{\mu }+\frac{1}{2}\partial _{\mu }\varphi \partial ^{\mu }\varphi +%
\frac{1}{2}GG  \label{9A} \\
&&+\varepsilon ^{\mu \nu \kappa \lambda }A_{\mu }\partial _{\nu
}A_{\kappa }p_{\lambda }+\frac{1}{4}\varepsilon ^{\mu \nu \kappa
\lambda }F_{\mu \nu }\varphi S_{\kappa \lambda },  \nonumber
\end{eqnarray}
where $\phi $ works as a Lagrange multiplier. Notice that the
auxiliary field $\eta _{\mu }$ must satisfy the equation
(\ref{7}). The equations (\ref {8}) and (\ref{9}) can be seen as
central charge transformations. In this
way, the supersymmetric transformations given explicitly in 4D will contain $%
x^{4}$ translations that are interpreted as central charge
transformations. Then, we notice that this extra dimension works
as a central charge for the superalgebra in $4D$. This results are
possible because we used the technique of dimensional reduction
\`{a} la Legendre. We can observe that the action of this
Lagrangian is independent of the $x^{4}$-translations, although
the fields depend on this coordinate. For that, the Lagrangian is
required to obey all the constraint imposed by the equation of
motion (\ref {6}) in 5 dimensions, to be invariant under $x^{4}$
translations.

\subsection{The chain of reductions $10D\rightarrow 5D\rightsquigarrow 4D$}

Using the same idea as worked out for $N=2$, we can obtain the
bosonic sector of the off-shell $N=4-$SUSY version of the model
(\ref{0}), but, to achieve this, we need to start off from a
$D=10$ Lagrangian. It can be done making the reduction
$D=10\rightarrow D=5$ through the ordinary technique and then
reducing $D=5\rightsquigarrow D=4$ through the Legendre
transformation technique. As we see in \cite{3ca}, the Lagrangian
treated in $10D$ can be written as:
\begin{equation}
{\cal L}_{10}=-\frac{1}{4}F_{\dot{\mu}\dot{\nu}}F^{\dot{\mu}\dot{\nu}}+\frac{%
1}{7!}\varepsilon ^{\dot{\mu}\dot{\nu}\dot{\kappa}\dot{\lambda}\dot{\rho}%
\dot{\sigma}\dot{\alpha}\dot{\beta}\dot{\gamma}\dot{\delta}}A_{\dot{\mu}%
}\partial _{\dot{\nu}}A_{\dot{\kappa}}T_{\dot{\lambda}\dot{\rho}\dot{\sigma}%
\dot{\alpha}\dot{\beta}\dot{\gamma}\dot{\delta}},
\end{equation}
where $\dot{\mu}=0,1,2,3,4,5,6,7,8,9.$

It is easier in the calculations to consider the dual of the
Lorentz violating term, although we will continue to show the
results under the conventional form for the Chern-Simons action.
The dimensional reduction
\`{a} la Scherk from $10\rightarrow 5$ of the space-like coordinates $%
x^{5},x^{6},x^{7},x^{8},x^{9}$ is carried out by considering no
dependence of the fields on these coordinates. This reduction
takes the form as follows:
\begin{eqnarray}
{\cal L}_{5} &=&-\frac{1}{4}F_{\hat{\mu}\hat{\nu}}F^{\hat{\mu}\hat{\nu}}+%
\frac{1}{2}\partial _{\hat{\mu}}\varphi ^{I}\partial
^{\hat{\mu}}\varphi
^{I}+\frac{1}{2}\varepsilon ^{\hat{\mu}\hat{\nu}\hat{\kappa}\hat{\lambda}%
\hat{\rho}}A_{\hat{\mu}}\partial _{\hat{\nu}}A_{\hat{\kappa}}R_{\hat{\lambda}%
\hat{\rho}}  \label{16} \\
&&+\frac{1}{3!}\varepsilon ^{\hat{\mu}\hat{\nu}\hat{\kappa}\hat{\lambda}\hat{%
\rho}}A_{\hat{\mu}}\partial _{\hat{\nu}}\varphi ^{I}T_{\hat{\kappa}\hat{%
\lambda}\hat{\rho}}^{I}+\frac{1}{4!}\varepsilon ^{\hat{\mu}\hat{\nu}\hat{%
\kappa}\hat{\lambda}\hat{\rho}}\varphi ^{I}\partial
_{\hat{\mu}}\varphi
^{J}T_{\hat{\nu}\hat{\kappa}\hat{\lambda}\hat{\rho}}^{IJ}.
\nonumber
\end{eqnarray}
where $\hat{\mu}=0,1,2,3,4$, the internal index $I=1,2,3,4,5$
and\thinspace
\thinspace $\varepsilon ^{\hat{\mu}\hat{\nu}\hat{\kappa}\hat{\lambda}\hat{%
\rho}56789}\equiv \varepsilon ^{\hat{\mu}\hat{\nu}\hat{\kappa}\hat{\lambda}%
\hat{\rho}}.$ The gauge fields were redefined as:
\[
(A_{\dot{\mu}})\rightarrow (A_{\hat{\mu}},\,\varphi ^{I}),
\]
and the background fields as
\[
(T_{\dot{\lambda}\dot{\rho}\dot{\sigma}\dot{\alpha}\dot{\beta}\dot{\gamma}%
\dot{\delta}})\rightarrow (R_{\hat{\lambda}\hat{\rho}},\,T_{\hat{\kappa}\hat{%
\lambda}\hat{\rho}}^{I},\,T_{\hat{\nu}\hat{\kappa}\hat{\lambda}\hat{\rho}%
}^{IJ}).
\]
We can notice that the Lagrangian in $5D$ has a vector and five
scalar gauge fields; one rank-2, five rank-3 (its dual is rank-2)
and ten rank-4 (its dual is a vector) background fields.

Now, we proceed performing the dimensional reduction of the
space-like $x^{4} $-coordinate by using the Legendre
transformation technique. This will bring about the auxiliary
fields that make the superalgebra closed off-shell. In this
technique, the $D=4$ Lagrangian is obtained through
\begin{equation}
{\cal L}_{4}={\cal L}_{5}-\frac{\partial {\cal L}_{5}}{\partial
\partial
_{4}A_{\hat{\nu}}}\frac{\partial A_{\hat{\nu}}}{\partial x^{4}}-\frac{%
\partial {\cal L}_{5}}{\partial \partial _{4}\varphi ^{I}}\frac{\partial
\varphi ^{I}}{\partial x^{4}}.  \label{17}
\end{equation}
Applying the (\ref{17}) for the Lagrangian (\ref{16}), we obtain:
\begin{eqnarray}
{\cal L}_{4} &=&-\frac{1}{4}F_{\mu \nu }F^{\mu \nu
}+\frac{1}{2}F_{4\mu }F^{4\mu }+\partial _{\mu }\phi F^{4\mu
}+\frac{1}{2}\partial _{\mu }\varphi ^{I}\partial ^{\mu }\varphi
^{I}-\frac{1}{2}\partial _{4}\varphi
^{I}\partial ^{4}\varphi ^{I}  \label{18} \\
&&+\varepsilon ^{\mu \nu \kappa \lambda }A_{\mu }\partial _{\nu
}A_{\kappa }p_{\lambda }+\varepsilon ^{\mu \nu \kappa \lambda
}A_{\mu }\partial _{\nu
}\phi R_{\kappa \lambda } \\
&&+\frac{1}{2}\varepsilon ^{\mu \nu \kappa \lambda }A_{\mu
}\partial _{\nu }\varphi ^{I}S_{\kappa \lambda }^{I}+\phi \partial
_{\mu }\varphi ^{I}t^{I\mu }+\varphi ^{I}\partial _{\mu }\varphi
^{J}t^{IJ\mu },  \nonumber
\end{eqnarray}
where $\varepsilon ^{\mu \nu \kappa \lambda }\equiv \varepsilon
^{\mu \nu \kappa \lambda 4}$, and we redefined the gauge fields as
\[
(A_{\hat{\mu}},\,\varphi ^{I})\rightsquigarrow (A_{\mu },\,\phi
,\varphi ^{I}),
\]
and the background fields as
\[
(R_{\hat{\lambda}\hat{\rho}},\,T_{\hat{\kappa}\hat{\lambda}\hat{\rho}%
}^{I},\,T_{\hat{\nu}\hat{\kappa}\hat{\lambda}\hat{\rho}}^{IJ})%
\rightsquigarrow (p_{\lambda },\,R_{\lambda \rho },\,S_{\kappa
\lambda }^{I},\,t^{I\mu },\,t^{IJ\mu },\,t^{IJ}).
\]
We notice that the theory in $4D$ has one vector and six scalar
gauge fields; sixteen vector (given by one vector and fifteen
duals of rank-3 tensors), ten scalars (given by ten duals of
rank-4 tensors) and six rank-2 tensor background fields.

The canonical momenta for the Lagrangian (\ref{16}) are given as
follows:
\begin{eqnarray*}
\pi (A_{\mu }) &=&\frac{\partial {\cal L}_{5}}{\partial \partial _{4}A_{\mu }%
}=-F^{4\mu }+\frac{1}{2}\varepsilon ^{\mu \nu \kappa \lambda
}A_{\nu
}R_{\kappa \lambda }, \\
\pi (\varphi ) &=&\frac{\partial {\cal L}_{5}}{\partial \partial
_{4}\varphi ^{I}}=\partial ^{4}\varphi ^{I}-A_{\mu }t^{I\mu
}+\varphi ^{J}t^{IJ}.
\end{eqnarray*}
We define the new auxiliary fields as
\[
F_{4\mu }\equiv \eta _{\mu },\,\,\,F_{4}^{\,\,\,\,\mu }=-F^{4\mu
}\equiv \eta ^{\mu },\,\,\,\,\,\partial _{4}\varphi ^{I}=-\partial
^{4}\varphi ^{I}\equiv G^{I}.
\]
This will be necessary to close the superalgebra off-shell. By
bringing these definitions into the Lagrangian (\ref{18}), we end
up with
\begin{eqnarray}
{\cal L}_{4} &=&-\frac{1}{4}F_{\mu \nu }F^{\mu \nu
}-\frac{1}{2}\eta _{\mu }\eta ^{\mu }-\eta ^{\mu }\partial _{\mu
}\phi +\frac{1}{2}\partial _{\mu }\varphi ^{I}\partial ^{\mu
}\varphi ^{I}+\frac{1}{2}G^{I}G^{I}+\varepsilon
^{\mu \nu \kappa \lambda }A_{\mu }\partial _{\nu }A_{\kappa }p_{\lambda } \\
&&+\varepsilon ^{\mu \nu \kappa \lambda }A_{\mu }\partial _{\nu
}\phi R_{\kappa \lambda }+\frac{1}{2}\varepsilon ^{\mu \nu \kappa
\lambda }A_{\mu }\partial _{\nu }\varphi ^{I}S_{\kappa \lambda
}^{I}+\phi \partial _{\mu }\varphi ^{I}t^{I\mu }+\varphi
^{I}\partial _{\mu }\varphi ^{J}t^{IJ\mu }. \nonumber
\end{eqnarray}
The equation of motion for the Lagrangian (\ref{16}) in terms of $A_{\hat{\nu%
}}$ is given by:
\begin{equation}
\partial _{\hat{\mu}}F^{\hat{\mu}\hat{\nu}}+\varepsilon ^{\hat{\nu}\hat{\mu}%
\hat{\kappa}\hat{\lambda}\hat{\rho}}F_{\hat{\mu}\hat{\kappa}}R_{\hat{\lambda}%
\hat{\rho}}+\frac{1}{3!}\varepsilon ^{\hat{\nu}\hat{\mu}\hat{\kappa}\hat{%
\lambda}\hat{\rho}}\partial _{\hat{\mu}}\varphi ^{I}T_{\hat{\kappa}\hat{%
\lambda}\hat{\rho}}^{I}=0.
\end{equation}
Considering $\hat{\nu}=4$, we have that
\begin{equation}
\partial _{\mu }\eta ^{\mu }-\frac{1}{2}\varepsilon ^{\mu \nu \kappa \lambda
}F_{\mu \nu }R_{\kappa \lambda }-\partial _{\mu }\varphi
^{I}t^{I\mu }=0, \label{18A}
\end{equation}
and considering $\hat{\nu}=\nu ,$ we have that
\begin{equation}
\partial _{4}\eta ^{\nu }=\partial _{\mu }F^{\mu \nu }+\frac{1}{2}%
\varepsilon ^{\nu \mu \kappa \lambda }F_{\mu \kappa }p_{\lambda }+\frac{1}{2}%
\varepsilon ^{\nu \mu \kappa \lambda }\partial _{\mu }\varphi
^{I}S_{\kappa \lambda }^{I}-\varepsilon ^{\nu \mu \kappa \lambda
}\eta _{\mu }R_{\kappa \lambda }+G^{I}t^{I\mu }.  \label{19}
\end{equation}
The equation of motion for $\varphi $ is presented as follows
\begin{equation}
\partial _{\hat{\mu}}\partial ^{\hat{\mu}}\varphi ^{I}-\frac{1}{3!}%
\varepsilon
^{\hat{\mu}\hat{\nu}\hat{\kappa}\hat{\lambda}\hat{\rho}}\partial
_{\hat{\mu}}A_{\hat{\nu}}T_{\hat{\kappa}\hat{\lambda}\hat{\rho}}^{I}-\frac{2%
}{4!}\varepsilon ^{\hat{\mu}\hat{\nu}\hat{\kappa}\hat{\lambda}\hat{\rho}%
}\partial _{\hat{\mu}}\varphi ^{J}T_{\hat{\nu}\hat{\kappa}\hat{\lambda}\hat{%
\rho}}^{IJ}=0.
\end{equation}
Taking the component $\hat{\mu}=4$, we arrive at
\begin{equation}
\partial _{4}G^{I}=\partial _{\mu }\partial ^{\mu }\varphi ^{I}-\frac{1}{4}%
\varepsilon ^{\mu \kappa \lambda \rho }F_{\mu \nu }S_{\lambda \rho
}^{I}-\eta _{\mu }t^{I\mu }-2\partial _{\mu }\varphi ^{J}t^{IJ\mu
}-2G^{J}t^{IJ}.  \label{19A}
\end{equation}

Notice that the constraint (\ref{7}) can directly simplify the
Lagrangian to
\begin{eqnarray}
{\cal L}_{4} &=&-\frac{1}{4}F_{\mu \nu }F^{\mu \nu
}-\frac{1}{2}\eta _{\mu }\eta ^{\mu }+\frac{1}{2}\partial _{\mu
}\varphi ^{I}\partial ^{\mu }\varphi
^{I}+\frac{1}{2}G^{I}G^{I}  \label{20} \\
&&+\varepsilon ^{\mu \nu \kappa \lambda }A_{\mu }\partial _{\nu
}A_{\kappa }p_{\lambda }+\frac{1}{4}\varepsilon ^{\mu \nu \kappa
\lambda }F_{\mu \nu }\varphi ^{I}S_{\kappa \lambda }^{I}+\varphi
^{I}\partial _{\mu }\varphi ^{J}t^{IJ\mu }.  \nonumber
\end{eqnarray}
As in the 6 to 4 reduction the field $\phi $ works as a Lagrange
multiplier. The auxiliary field $\eta _{\mu }$ must satisfy the
equation (\ref{18A}) and the $x^{4}$-translations
(\ref{19},\ref{19A}) correspond the central charge transformations
in the $N=4$ superalgebra for $D=4$.

\section{The off-shell $N=2$-SUSY version of the Abelian gauge model with
Lorentz-breaking term}

The on-shell version of the $N=2$ supersymmetric extension of the
Lorentz breaking term can be found in \cite{3ca}, where use has
been made of the dimensional reduction (\`{a} la Scherk). In the
present work, we are interested in the attainment of the off-shell
$N=2$ SUSY version of the Abelian gauge model with
Lorentz-breaking term. In this way, we consider that the bosonic
sector for $N=1$ in $6D$ is the same of the bosonic sector for
$N=2$ in $4D.$ In order to build up the supersymmetrization
off-shell, it is necessary to reduce one of the coordinates using
the reduction \`{a} la Legendre. It permits the appearance of
auxiliary fields that make the algebra closed off-shell.

In the previous section, we have made the dimensional reduction $%
6\rightarrow 5$ \`{a} la Scherk and $5\rightsquigarrow 4$ \`{a} la
Legendre and obtained the bosonic Lagrangian (\ref{9A}) in $4D$.
We can consider the vector field of the theory as a gradient of a
scalar ($p_{\mu }=\partial _{\mu }s$), then the Lagrangian is
given as follows:
\begin{eqnarray}
{\cal L}_{4} &=&-\frac{1}{4}F_{\mu \nu }F^{\mu \nu
}-\frac{1}{2}\eta _{\mu }\eta ^{\mu }+\frac{1}{2}\partial _{\mu
}\varphi _{1}\partial ^{\mu }\varphi
_{1}+\frac{1}{2}G_{1}G_{1} \\
&&+\varepsilon ^{\mu \nu \kappa \lambda }A_{\mu }\partial _{\nu
}A_{\kappa }\partial _{\lambda }s_{1}+\frac{1}{4}\varepsilon ^{\mu
\nu \kappa \lambda }F_{\mu \nu }\varphi _{1}S_{\kappa \lambda }.
\nonumber
\end{eqnarray}
Notice that we redefined the real fields $\varphi ,\,s$ and $G$.
It is necessary in order to accommodate these bosonic component
fields in the chiral superfields. For that, we have to define the
complex fields:
\begin{eqnarray*}
\varphi &=&\varphi _{1}+i\varphi _{2} \\
s &=&s_{2}+is_{1}, \\
G &=&G_{1}+iG_{2}.
\end{eqnarray*}
Observe that we introduced more one gauge, one background and one
auxiliary fields in the model to build up complex scalar fields.
Once we has the bosonic sector of the $N=2$-SUSY theory, we
proceed the supersymmetrization using the superfield formulation
in a $N=1$ superspace with supercoordinates $(x^{\mu },\theta
^{a},\bar{\theta}_{\dot{a}}).$ The conventions used are the same
as given in ref. \cite{3ca}.

We define a vector superfields $V$ in the Wess-Zumino gauge
containing the gauge field $A_{\mu }$ as:
\begin{equation}
V=\theta \sigma ^{\mu }\bar{\theta}A_{\mu }+\theta ^{2}\bar{\theta}\bar{%
\lambda}+\bar{\theta}^{2}\theta \lambda +\theta
^{2}\bar{\theta}^{2}D \label{9C}
\end{equation}
which fulfill the reality constraint $V=V^{\dagger }.$ The Abelian
field-strength superfield is given by:
\begin{equation}
W_{a}=-\frac{1}{4}\bar{D}^{2}D_{a}V_{WZ},\,\,\,\,\,\,\,\,\,\,\bar{W}_{\dot{a}%
}=-\frac{1}{4}D^{2}\bar{D}_{\dot{a}}V_{WZ},
\end{equation}
having the chirality condition: $\bar{D}W=D\bar{W}=0$ and $DW=\bar{D}\bar{W}%
. $

The vector superfield that has the auxiliary field $\eta ^{\mu }$
must not be gauge invariant, then, it is necessary to define this
in the complete form:
\begin{equation}
U=u+\theta \alpha +\bar{\theta}\bar{\alpha}+\theta ^{2}M+\bar{\theta}%
^{2}M^{*}+\theta \sigma ^{\mu }\bar{\theta}\eta _{\mu }+\theta ^{2}\bar{%
\theta}\bar{\beta}+\bar{\theta}^{2}\theta \beta +\theta ^{2}\bar{\theta}%
^{2}E,  \label{9D}
\end{equation}
obeying the reality constraint, $U=U^{\dagger }.$

The scalar superfields that accommodate the gauge field $\varphi
,\,\varphi ^{*}$ and the auxiliary fields $G,G^{*}$ are:
\begin{equation}
\Phi =\varphi +i\theta \sigma ^{\mu }\bar{\theta}\partial _{\mu }\varphi -%
\frac{1}{4}\theta ^{2}\bar{\theta}^{2}\Box \varphi +\sqrt{2}\theta \psi -%
\frac{i}{\sqrt{2}}\theta ^{2}\partial _{\mu }\psi \sigma ^{\mu }\bar{\theta}%
+\theta ^{2}G,
\end{equation}
\begin{equation}
\bar{\Phi}=\varphi ^{*}-i\theta \sigma ^{\mu }\bar{\theta}\partial
_{\mu
}\varphi ^{*}-\frac{1}{4}\theta ^{2}\bar{\theta}^{2}\Box \varphi ^{*}+\sqrt{2%
}\bar{\theta}\bar{\psi}+\frac{i}{\sqrt{2}}\bar{\theta}^{2}\theta
\sigma ^{\mu }\partial _{\mu }\bar{\psi}+\bar{\theta}^{2}G^{*}.
\end{equation}
The scalar superfields that accommodate the background fields
$s,\,s^{*}$ and their superpartners are written, respectively as
\begin{equation}
S=s+i\theta \sigma ^{\mu }\bar{\theta}\partial _{\mu
}s-\frac{1}{4}\theta ^{2}\bar{\theta}^{2}\Box s+\sqrt{2}\theta \xi
-\frac{i}{\sqrt{2}}\theta ^{2}\partial _{\mu }\xi \sigma ^{\mu
}\bar{\theta}+\theta ^{2}h,  \label{10}
\end{equation}
\begin{equation}
\bar{S}=s^{*}-i\theta \sigma ^{\mu }\bar{\theta}\partial _{\mu }s^{*}-\frac{1%
}{4}\theta ^{2}\bar{\theta}^{2}\Box s^{*}+\sqrt{2}\bar{\theta}\bar{\xi}+%
\frac{i}{\sqrt{2}}\bar{\theta}^{2}\theta \sigma ^{\mu }\partial _{\mu }\bar{%
\xi}+\bar{\theta}^{2}h^{*},  \label{10a}
\end{equation}
which satisfy the chiral condition: $\bar{D}\Phi =D\bar{\Phi}=\bar{D}S=D\bar{%
S}=0.$ Observe that we introduced more one gauge, one auxiliary
and one background fields in the model to build up complex scalar
fields. It was necessary in order to accommodate the fields in the
chiral superfields.

The spinor superfields that contain $S_{\mu \nu }$, their dual
fields and their superpartners are written as
\begin{eqnarray}
\Sigma _{a} &=&\tau _{a}+\theta ^{b}(\varepsilon _{ba}\rho +\sigma
_{ba}^{\mu \nu }S_{\mu \nu })+\theta ^{2}F_{a}+i\theta \sigma ^{\mu }\bar{%
\theta}\partial _{\mu }\tau _{a}  \label{12} \\
&&+i\theta \sigma ^{\mu }\bar{\theta}\theta ^{b}\partial _{\mu
}(\varepsilon
_{ba}\rho +\sigma _{ba}^{\mu \nu }S_{\mu \nu })-\frac{1}{4}\theta ^{2}\bar{%
\theta}^{2}\square \tau _{a},  \nonumber
\end{eqnarray}
\begin{eqnarray}
\bar{\Sigma}_{\dot{a}} &=&\bar{\tau}_{\dot{a}}+\bar{\theta}_{\dot{b}%
}(-\varepsilon _{\,\,\dot{a}}^{\dot{b}}\rho ^{*}-\bar{\sigma}%
_{\,\,\,\,\,\,\,\,\,\,\,\dot{a}}^{\mu \nu \dot{b}}S_{\mu \nu }^{*})+\bar{%
\theta}^{2}\bar{F}_{\dot{a}}-i\theta \sigma ^{\mu
}\bar{\theta}\partial
_{\mu }\bar{\tau}_{\dot{a}} \\
&&-i\theta \sigma ^{\mu }\bar{\theta}\theta _{\dot{b}}\partial
_{\mu
}(-\varepsilon _{\,\,\,\dot{a}}^{\dot{b}}\rho ^{*}-\bar{\sigma}%
_{\,\,\,\,\,\,\,\,\,\,\,\dot{a}}^{\mu \nu \dot{b}}S_{\mu \nu }^{*})-\frac{1}{%
4}\theta ^{2}\bar{\theta}^{2}\square \bar{\tau}_{\dot{a}},
\nonumber
\end{eqnarray}
that are also chiral $\bar{D}_{\dot{b}}\Sigma _{a}=D_{b}\bar{\Sigma}_{\dot{a}%
}=0.$

Now, we are interested in building up the off-shell $N=2$
supersymmetric version of the Lagrangian (\ref{0}) using the
$N=1$-superfield formalism. As already written down in the
previous section, the bosonic sector for this Lagrangian is given
by (\ref{9A}). In this way, the next step is to look for a
supersymmetric model in term of superfields that contain this
bosonic sector and its correspondent fermionic sector. First, it
is useful to quote the mass dimensions of the superfields given
previously:
\begin{eqnarray*}
\lbrack V] &=&0,\,\,\,\,\,\,\,\,[W_{a}]=[\bar{W}_{\dot{a}}]=+\frac{3}{2}%
,\,\,\,\,\,\,\,\,[\Phi ]=[\bar{\Phi}]=+1, \\
\lbrack U] &=&[\bar{U}]=+1,\,\,\,\,\,\,\,[S]=[\bar{S}]=0,\,\,\,\,\,\,\,\,[%
\Sigma _{a}]=[\bar{\Sigma}_{\dot{a}}]=+1.
\end{eqnarray*}
Based on the dimensionalities, and by analyzing the bosonic
Lagrangian (\ref {9A}), we propose the following supersymmetric
action ${\cal S}_{br}$:

\begin{eqnarray}
{\cal S}_{br} &=&\int d^{4}xd^{2}\theta d^{2}\bar{\theta}[\frac{1}{4}%
W^{\alpha }W_{\alpha }\delta (\bar{\theta}^{2})+\frac{1}{4}\bar{W}_{\dot{%
\alpha}}\bar{W}^{\dot{\alpha}}\delta (\theta
^{2})+\frac{1}{2}\bar{\Phi}\Phi
-UU  \nonumber \\
&&+\frac{1}{2}W^{a}(D_{a}V)S+\frac{1}{2}\bar{W}_{\dot{a}}(\bar{D}^{\dot{a}}V)%
\bar{S} \\
&&+\frac{i}{4}\delta (\bar{\theta})W^{a}(\Phi +\bar{\Phi})\Sigma _{a}-\frac{i%
}{4}\delta (\theta )\bar{W}_{\dot{a}}(\Phi +\bar{\Phi})\bar{\Sigma}^{\dot{a}%
}].  \nonumber
\end{eqnarray}
This Lagrangian in its component-field reads as below:
\begin{eqnarray}
{\cal L}_{br} &=&-\frac{1}{4}F_{\mu \nu }F^{\mu \nu }-i\lambda
\sigma ^{\mu }\partial _{\mu
}\bar{\lambda}+D^{2}+D^{*2}+\frac{1}{2}\partial _{\mu }\varphi
\partial ^{\mu }\varphi ^{*}-\frac{i}{2}\psi \sigma ^{\mu
}\partial
_{\mu }\bar{\psi}+\frac{1}{2}GG^{*}  \nonumber \\
&&-\frac{1}{2}\eta ^{\mu }\eta _{\mu }+\alpha \beta +\bar{\alpha}\bar{\beta}%
-MM^{*}-2uE  \nonumber \\
&&+\frac{i}{4}\partial _{\mu }(s-s^{*})\varepsilon ^{\mu \kappa
\lambda \nu }F_{\kappa \lambda }A_{\nu
}-\frac{1}{4}(s+s^{*})F_{\mu \nu }F^{\mu \nu
}+2D^{2}(s+s^{*})  \nonumber \\
&&-is\lambda \sigma ^{\mu }\partial _{\mu }\bar{\lambda}-is^{*}\bar{\lambda}%
\bar{\sigma}^{\mu }\partial _{\mu }\lambda
-\frac{1}{\sqrt{2}}\lambda \sigma ^{\mu \nu }F_{\mu \nu }\xi
+\frac{1}{\sqrt{2}}\bar{\lambda}\bar{\sigma}^{\mu
\nu }F_{\mu \nu }\bar{\xi}  \nonumber \\
&&+\frac{1}{2}\lambda \lambda h+\frac{1}{2}\bar{\lambda}\bar{\lambda}h^{*}-%
\sqrt{2}\lambda \xi D-\sqrt{2}\bar{\lambda}\bar{\xi}D  \nonumber \\
&&\frac{1}{16}\varepsilon ^{\mu \nu \kappa \lambda }F_{\mu \nu
}(\varphi
+\varphi ^{*})(R_{\kappa \lambda }+R_{\kappa \lambda }^{*})+\frac{i}{8}%
F^{\mu \nu }(R_{\mu \nu }-R_{\mu \nu }^{*})(\varphi +\varphi ^{*})
\label{14} \\
&&-\frac{i\sqrt{2}}{8}\tau \sigma ^{\mu \nu }\psi F_{\mu \nu }-\frac{i\sqrt{2%
}}{8}\bar{\tau}\bar{\sigma}^{\mu \nu }\bar{\psi}F_{\mu \nu
}+\frac{1}{4}\tau \sigma ^{\mu }\partial _{\mu
}\bar{\lambda}(\varphi +\varphi ^{*})  \nonumber
\\
&&-\frac{1}{4}\bar{\tau}\bar{\sigma}^{\mu }\partial _{\mu }\lambda
(\varphi +\varphi ^{*})+\frac{i\sqrt{2}}{4}\psi \sigma ^{\mu \nu
}B_{\mu \nu }\lambda
+\frac{i\sqrt{2}}{4}\bar{\psi}\bar{\sigma}^{\mu \nu }B_{\mu \nu }^{*}\bar{%
\lambda}  \nonumber \\
&&-\frac{i}{2}D(\varphi +\varphi ^{*})\rho
+\frac{i}{2}D^{*}(\varphi
+\varphi ^{*})\rho ^{*}  \nonumber \\
&&+\frac{i\sqrt{2}}{8}\lambda \psi \rho -\frac{i\sqrt{2}}{8}\bar{\lambda}%
\bar{\psi}\rho ^{*}-\frac{i\sqrt{2}}{4}D\psi \tau +\frac{i\sqrt{2}}{4}D^{*}%
\bar{\psi}\bar{\tau}  \nonumber \\
&&+\frac{i}{4}f\lambda \tau -\frac{i}{4}f^{*}\bar{\lambda}\bar{\tau}+\frac{i%
}{4}(\varphi +\varphi ^{*})\lambda F-\frac{i}{4}(\varphi +\varphi ^{*})\bar{%
\lambda}\bar{F}.  \nonumber
\end{eqnarray}
This Lagrangian is invariant under the $N=2$ superalgebra in this
$4D$. When the superalgebra in $5D$ has the index 4 explicited we
obtain the superalgebra in $4D$ that acquire central charge
transformations that correspond the $x^{4}$ transformations as we
obtained in (\ref{8},\thinspace \ref{9}). This is given as
follows:
\begin{eqnarray}
\delta _{Z}\eta ^{\nu } &=&\partial _{\mu }F^{\mu \nu }+\frac{1}{2}%
\varepsilon ^{\nu \mu \kappa \lambda }F_{\mu \kappa }p_{\lambda
}-\varepsilon ^{\nu \mu \kappa \lambda }\eta _{\mu }R_{\kappa \lambda } \\
&&+\frac{1}{2}\varepsilon ^{\nu \mu \kappa \lambda }\partial _{\mu
}\varphi
S_{\kappa \lambda }-Gt^{\mu }+fermionic\,\,SUSY\,\,partners,  \nonumber \\
\delta _{Z}G &=&\partial _{\mu }\partial ^{\mu }\varphi -\frac{1}{4}%
\varepsilon ^{\mu \kappa \lambda \rho }F_{\mu \nu }S_{\lambda \rho
}-\eta _{\mu }t^{\mu }+fermionic\,\,SUSY\,\,partners.
\end{eqnarray}
These central charge transformations above include only the
bosonic sector, because we do not start off with the fermions in
$6D$. Once we have the reduced fermionic term (in $4D$), we are
able to restore the fermions of the
original theory in $6D$. With that, our central charge transformations in $%
4D $ will naturally display the fermionic fields along with the
bosonic degrees of freedom. This shall be presented in a
forthcoming work.

The fifth dimension is interpreted as the central charge and it is
consequence of the necessity of the obedience of the equation of motion in $%
5D$.

We can notice that this off-shell Lagrangian is simpler than the
on-shell one obtained in the in ref. \cite{3ca}. This is so by
virtue of the constraint (\ref{7}) imposed by the equation of
motion in 5 dimensions. These results are possible due to the
utility of the Legendre transformation technique to make the
dimensional reduction of one of the space-like coordinates.

In the Lagrangian (\ref{14}), we can notice the Maxwell term and
the term proposed by Jackiw in \cite{1} and also the $N=1$
supersymmetric generalization. We also see that this Lagrangian
has the bosonic sector (\ref {9A}), the fermionic sector and the
auxiliary fields presented in the superfields.

\section{The off-shell $N=4$-SUSY version of the Abelian gauge model with
Lorentz-breaking term}

In order to build up the off-shell $N=4$ supersymmetric version of
the Abelian gauge model with Lorentz violating term, we proceed
similar to the last section, but now, we start with the bosonic
Lagrangian (\ref{20}) obtained through the reduction
$10\rightarrow 5$ and $5\rightsquigarrow 4$. As in the last
section, we consider the vector fields as gradient of scalars
($p_{\mu }=\partial _{\mu }s,\,\,\,t^{IJ\mu }=\partial ^{\mu
}u^{IJ}$), then the bosonic Lagrangian is given as follows:
\begin{eqnarray}
{\cal L}_{4} &=&-\frac{1}{4}F_{\mu \nu }F^{\mu \nu
}-\frac{1}{2}\eta _{\mu }\eta ^{\mu }+\frac{1}{2}\partial _{\mu
}\varphi _{1}^{I}\partial ^{\mu
}\varphi _{1}^{I}+\frac{1}{2}G_{1}^{I}G_{1}^{I}  \label{30} \\
&&+\varepsilon ^{\mu \nu \kappa \lambda }A_{\mu }\partial _{\nu
}A_{\kappa }\partial _{\lambda }s_{2}+\frac{1}{4}\varepsilon ^{\mu
\nu \kappa \lambda }F_{\mu \nu }\varphi _{1}^{I}S_{\kappa \lambda
}^{I}+\varphi _{1}^{I}\partial _{\mu }\varphi _{1}^{J}\partial
^{\mu }u_{1}^{IJ}. \nonumber
\end{eqnarray}
As it was done in the previous section, we redefined the real fields $%
\varphi ^{I}$, $s$ and $G^{I}$ in order to accommodate these
bosonic component fields in the chiral superfields. We define the
complex fields:
\begin{eqnarray*}
\varphi ^{I} &=&\varphi _{1}^{I}+i\varphi _{2}^{I}, \\
s &=&s_{2}+is_{1}, \\
G^{I} &=&G_{1}^{I}+iG_{2}^{I}, \\
u^{IJ} &=&u_{1}^{IJ}+iu_{2}^{IJ}
\end{eqnarray*}
We proceed the supersymmetrization building up the
$N=1-$superfield extension of the Lagrangian (\ref{30}). The
superfields that accommodates the gauge field $A_{\mu }$, the
background scalar fields $s,\,s^{*}$ and the auxiliary vector
field $\eta ^{\mu }$ with theirs superpartners is the same of the
defined in the last section, given by (\ref{9C}, \ref{10},
\ref{10a}, \ref{9D}) respectively. The scalar superfields that
accommodate the five real scalar gauge fields $\varphi _{1}^{I}$,
and introducing more five real
fields $\varphi _{2}^{I}$ that does not appears in the Lagrangian (\ref{20}%
), are given, respectively as
\begin{equation}
\Phi ^{I}=\varphi ^{I}+i\theta \sigma ^{\mu }\bar{\theta}\partial
_{\mu
}\varphi ^{I}-\frac{1}{4}\theta ^{2}\bar{\theta}^{2}\Box \varphi ^{I}+\sqrt{2%
}\theta \psi ^{I}-\frac{i}{\sqrt{2}}\theta ^{2}\partial _{\mu
}\psi ^{I}\sigma ^{\mu }\bar{\theta}+\theta ^{2}G^{I},
\end{equation}
\begin{equation}
\bar{\Phi}^{I}=\varphi ^{*I}-i\theta \sigma ^{\mu
}\bar{\theta}\partial _{\mu }\varphi ^{*I}-\frac{1}{4}\theta
^{2}\bar{\theta}^{2}\Box \varphi
^{*I}+\sqrt{2}\bar{\theta}\bar{\psi}^{I}+\frac{i}{\sqrt{2}}\bar{\theta}%
^{2}\theta \sigma ^{\mu }\partial _{\mu }\bar{\psi}^{I}+\bar{\theta}%
^{2}G^{*I},
\end{equation}
which satisfy the chiral condition: $\bar{D}\Phi
^{I}=D\bar{\Phi}^{I}=0.$ The chiral superfields that accommodate
$u^{IJ}$ and $u^{*IJ}$ are
\begin{equation}
R^{IJ}=u^{IJ}+i\theta \sigma ^{\mu }\bar{\theta}\partial _{\mu }u^{IJ}-\frac{%
1}{4}\theta ^{2}\bar{\theta}^{2}\Box u^{IJ}+\sqrt{2}\theta \zeta ^{IJ}-\frac{%
i}{\sqrt{2}}\theta ^{2}\partial _{\mu }\zeta ^{IJ}\sigma ^{\mu }\bar{\theta}%
+\theta ^{2}g^{I},
\end{equation}
\begin{equation}
\bar{R}^{IJ}=u^{*IJ}-i\theta \sigma ^{\mu }\bar{\theta}\partial
_{\mu
}u^{*IJ}-\frac{1}{4}\theta ^{2}\bar{\theta}^{2}\Box u^{*IJ}+\sqrt{2}\bar{%
\theta}\bar{\zeta}^{IJ}+\frac{i}{\sqrt{2}}\bar{\theta}^{2}\theta
\sigma ^{\mu }\partial _{\mu
}\bar{\zeta}^{IJ}+\bar{\theta}^{2}g^{*I},
\end{equation}

The spinor superfields that contain $S_{\mu \nu }^{I}$, their dual
fields and their superpartners are written as
\begin{eqnarray}
\Sigma _{a}^{I} &=&\tau _{a}^{I}+\theta ^{b}(\varepsilon _{ba}\rho
^{I}+\sigma _{ba}^{\mu \nu }S_{\mu \nu }^{I})+\theta
^{2}F_{a}^{I}+i\theta
\sigma ^{\mu }\bar{\theta}\partial _{\mu }\tau _{a}^{I} \\
&&+i\theta \sigma ^{\mu }\bar{\theta}\theta ^{b}\partial _{\mu
}(\varepsilon _{ba}\rho ^{I}+\sigma _{ba}^{\mu \nu }S_{\mu \nu
}^{I})-\frac{1}{4}\theta ^{2}\bar{\theta}^{2}\square \tau
_{a}^{I},  \nonumber
\end{eqnarray}
\begin{eqnarray}
\bar{\Sigma}_{\dot{a}}^{I} &=&\bar{\tau}_{\dot{a}}^{I}+\bar{\theta}_{\dot{b}%
}(-\varepsilon _{\,\,\dot{a}}^{\dot{b}}\rho ^{*I}-\bar{\sigma}%
_{\,\,\,\,\,\,\,\,\,\,\,\dot{a}}^{\mu \nu \dot{b}}S_{\mu \nu }^{*I})+\bar{%
\theta}^{2}\bar{F}_{\dot{a}}^{I}-i\theta \sigma ^{\mu
}\bar{\theta}\partial
_{\mu }\bar{\tau}_{\dot{a}}^{I} \\
&&-i\theta \sigma ^{\mu }\bar{\theta}\theta _{\dot{b}}\partial
_{\mu
}(-\varepsilon _{\,\,\,\dot{a}}^{\dot{b}}\rho ^{*I}-\bar{\sigma}%
_{\,\,\,\,\,\,\,\,\,\,\,\dot{a}}^{\mu \nu \dot{b}}S_{\mu \nu }^{*I})-\frac{1%
}{4}\theta ^{2}\bar{\theta}^{2}\square \bar{\tau}_{\dot{a}}^{I},
\nonumber
\end{eqnarray}
that are also chiral $\bar{D}_{\dot{b}}\Sigma _{a}^{I}=D_{b}\bar{\Sigma}_{%
\dot{a}}^{I}=0.$

Now, we are interested in building up the off-shell $N=4$
supersymmetric version of the Lagrangian (\ref{0}). The bosonic
sector for this Lagrangian is given by (\ref{20}). Now, we look
for a supersymmetric model in terms of superfields. The dimensions
of the new superfields are given as:
\[
\lbrack \Phi
^{I}]=[\bar{\Phi}^{I}]=+1,\,\,\,\,\,\,\,\,\,\,\,\,\,\,\,[\Sigma
_{a}^{I}]=[\bar{\Sigma}_{\dot{a}}^{I}]=+1,\,\,\,\,\,\,\,\,\,\,\,\,\,\,%
\,[R^{IJ}]=[\bar{R}^{IJ}]=0.
\]
Based on the dimensionalities, and by analyzing the bosonic
Lagrangian (\ref {20}), we propose the following supersymmetric
action, $S_{br}$:

\begin{eqnarray}
{\cal S}_{br} &=&\int d^{4}xd^{2}\theta d^{2}\bar{\theta}[\frac{1}{4}\bar{W}%
_{\dot{\alpha}}\bar{W}^{\dot{\alpha}}\delta (\theta ^{2})+\frac{1}{4}%
W^{\alpha }W_{\alpha }\delta (\bar{\theta}^{2})+\frac{1}{2}\bar{\Phi}%
^{I}\Phi ^{I}  \nonumber \\
&&-UU+\frac{1}{2}W^{a}(D_{a}V)S+\frac{1}{2}\bar{W}_{\dot{a}}(\bar{D}^{\dot{a}%
}V)\bar{S} \\
&&+\frac{i}{4}\delta (\bar{\theta})W^{a}(\Phi
^{I}+\bar{\Phi}^{I})\Sigma
_{a}^{I}-\frac{i}{4}\delta (\theta )\bar{W}_{\dot{a}}(\Phi ^{I}+\bar{\Phi}%
^{I})\bar{\Sigma}^{I\dot{a}}]  \nonumber \\
&&+\frac{1}{2}\Phi ^{I}\bar{\Phi}^{J}(R^{IJ}-\bar{R}^{IJ}).
\end{eqnarray}
This Lagrangian in its component-field version reads as below:
\begin{eqnarray}
{\cal L}_{br} &=&-\frac{1}{4}F_{\mu \nu }F^{\mu \nu }-i\lambda
\sigma ^{\mu }\partial _{\mu
}\bar{\lambda}+D^{2}+D^{*2}+\frac{1}{2}\partial _{\mu }\varphi
^{I}\partial ^{\mu }\varphi ^{*I}-\frac{i}{2}\psi ^{I}\sigma ^{\mu
}\partial _{\mu }\bar{\psi}^{I}+\frac{1}{2}G^{I}G^{*I}  \nonumber \\
&&-\frac{1}{2}\eta ^{\mu }\eta _{\mu }+\alpha \beta +\bar{\alpha}\bar{\beta}%
-MM^{*}-2uE  \nonumber \\
&&+\frac{i}{4}\partial _{\mu }(s-s^{*})\varepsilon ^{\mu \kappa
\lambda \nu }F_{\kappa \lambda }A_{\nu
}-\frac{1}{4}(s+s^{*})F_{\mu \nu }F^{\mu \nu
}+2D^{2}(s+s^{*})  \nonumber \\
&&-is\lambda \sigma ^{\mu }\partial _{\mu }\bar{\lambda}-is^{*}\bar{\lambda}%
\bar{\sigma}^{\mu }\partial _{\mu }\lambda
-\frac{1}{\sqrt{2}}\lambda \sigma ^{\mu \nu }F_{\mu \nu }\xi
+\frac{1}{\sqrt{2}}\bar{\lambda}\bar{\sigma}^{\mu
\nu }F_{\mu \nu }\bar{\xi}  \nonumber \\
&&+\frac{1}{2}\lambda \lambda h+\frac{1}{2}\bar{\lambda}\bar{\lambda}h^{*}-%
\sqrt{2}\lambda \xi D-\sqrt{2}\bar{\lambda}\bar{\xi}D  \nonumber \\
&&\frac{1}{16}\varepsilon ^{\mu \nu \kappa \lambda }F_{\mu \nu
}(\varphi
^{I}+\varphi ^{*I})(S_{\kappa \lambda }^{I}+S_{\kappa \lambda }^{*I})+\frac{i%
}{8}F^{\mu \nu }(S_{\mu \nu }^{I}-S_{\mu \nu }^{*I})(\varphi
^{I}+\varphi
^{*I})  \nonumber \\
&&-\frac{i\sqrt{2}}{8}\tau ^{I}\sigma ^{\mu \nu }\psi ^{I}F_{\mu \nu }-\frac{%
i\sqrt{2}}{8}\bar{\tau}^{I}\bar{\sigma}^{\mu \nu }\bar{\psi}^{I}F_{\mu \nu }+%
\frac{1}{4}\tau ^{I}\sigma ^{\mu }\partial _{\mu
}\bar{\lambda}(\varphi
^{I}+\varphi ^{*I})  \label{21} \\
&&-\frac{1}{4}\bar{\tau}^{I}\bar{\sigma}^{\mu }\partial _{\mu
}\lambda (\varphi ^{I}+\varphi ^{*I})+\frac{i\sqrt{2}}{4}\psi
^{I}\sigma ^{\mu \nu }S_{\mu \nu }^{I}\lambda
+\frac{i\sqrt{2}}{4}\bar{\psi}^{I}\bar{\sigma}^{\mu
\nu }S_{\mu \nu }^{I*}\bar{\lambda}^{I}  \nonumber \\
&&-\frac{i}{2}D(\varphi ^{I}+\varphi ^{*I})\rho ^{I}+\frac{i}{2}%
D^{*}(\varphi ^{I}+\varphi ^{*I})\rho ^{*I}  \nonumber \\
&&+\frac{i\sqrt{2}}{8}\lambda \psi ^{I}\rho ^{I}-\frac{i\sqrt{2}}{8}\bar{%
\lambda}\bar{\psi}^{I}\rho ^{*I}-\frac{i\sqrt{2}}{4}D\psi ^{I}\tau ^{I}+%
\frac{i\sqrt{2}}{4}D^{*}\bar{\psi}^{I}\bar{\tau}^{I}  \nonumber \\
&&+\frac{i}{4}f^{I}\lambda \tau ^{I}-\frac{i}{4}f^{*I}\bar{\lambda}\bar{\tau}%
^{I}+\frac{i}{4}(\varphi ^{I}+\varphi ^{*I})\lambda F^{I}-\frac{i}{4}%
(\varphi ^{I}+\varphi ^{*I})\bar{\lambda}\bar{F}^{I}  \nonumber \\
&&+\frac{1}{4}\varphi ^{I}\partial _{\mu }\varphi ^{*J}\partial
^{\mu }(u^{IJ}+u^{*IJ})-\frac{1}{4}\varphi ^{*J}\partial _{\mu
}\varphi
^{I}\partial ^{\mu }(u^{IJ}+u^{*IJ})  \nonumber \\
&&+\frac{1}{2}\partial ^{\mu }\varphi ^{I}\partial _{\mu }\varphi
^{*J}(u^{IJ}-u^{*IJ})-\frac{1}{4}\varphi ^{I}\varphi ^{*J}\square
(u^{IJ}-u^{*IJ})-\frac{i}{2}\psi ^{I}\sigma ^{\mu }\partial _{\mu }\bar{\psi}%
^{J}(u^{IJ}-u^{*IJ})  \nonumber \\
&&+\frac{1}{2}G^{I}G^{*J}(u^{IJ}-u^{*IJ})+\frac{i}{2}\psi ^{I}\sigma ^{\mu }%
\bar{\psi}^{J}\partial _{\mu }u^{*IJ}  \nonumber \\
&&-\frac{i}{2}\varphi ^{I}\zeta ^{IJ}\sigma ^{\mu }\partial _{\mu }\bar{\psi}%
^{J}+\frac{i}{2}\varphi ^{*J}\psi ^{I}\sigma ^{\mu }\partial _{\mu }\bar{%
\zeta}^{IJ}+\frac{i}{2}\psi ^{I}\sigma ^{\mu
}\bar{\zeta}^{IJ}\partial _{\mu
}\varphi ^{*J}  \nonumber \\
&&+\frac{1}{2}\varphi ^{I}G^{*J}g^{IJ}-\frac{1}{2}G^{I}\varphi ^{*J}g^{*IJ}-%
\frac{1}{2}G^{*J}\psi ^{I}\zeta ^{IJ}+\frac{1}{2}G^{I}\bar{\psi}^{J}\bar{%
\zeta}^{IJ}.  \nonumber
\end{eqnarray}
This $N=4$ Lagrangian contains the $N=1$ and $N=2$ terms. The
$x^{4}$ transformations given by (\ref{19},\ref{19A}) works as
central charge transformations in the superalgebra. These are
given by
\begin{eqnarray}
\delta _{Z}\eta ^{\nu } &=&\partial _{\mu }F^{\mu \nu }+\frac{1}{2}%
\varepsilon ^{\nu \mu \kappa \lambda }F_{\mu \kappa }\partial _{\lambda }s+%
\frac{1}{2}\varepsilon ^{\nu \mu \kappa \lambda }\partial _{\mu
}\varphi
^{I}S_{\kappa \lambda }^{I}  \nonumber \\
&&-\varepsilon ^{\nu \mu \kappa \lambda }\eta _{\mu }R_{\kappa
\lambda
}+G^{I}\partial ^{\mu }u^{I}+fermionic\,\,SUSY\,\,partners, \\
\delta _{Z}G^{I} &=&\partial _{\mu }\partial ^{\mu }\varphi ^{I}-\frac{1}{4}%
\varepsilon ^{\mu \kappa \lambda \rho }F_{\mu \nu }S_{\lambda \rho
}^{I}-\eta _{\mu }t^{I\mu }  \nonumber \\
&&-2\partial _{\mu }\varphi ^{J}\partial ^{\mu
}u^{IJ}-2G^{J}t^{IJ}+fermionic\,\,SUSY\,\,partners.
\end{eqnarray}
As already mentioned in the previous section, the fermionic terms
of the central charge transformations shall appear in a
forthcoming paper.

\section{Concluding Remarks and Comments}

In this work, we carried out the $N=2$ and $N=4$ supersymmetric
generalizations realized off-shell for the Abelian gauge model
with a
Chern-Simons Lorentz violating term starting with the bosonic sector of a $%
N=1$ version of this theory on $D=6$ and $D=4$ respectively, both
realized on-shell. Then, the dimensional reduction to $D=4$
yielded the bosonic sectors for the $N=2$ and $N=4$-SUSY versions
for this model. Once we had the bosonic sector, we could make the
supersymmetric extension using the superfields formalism in $N=1$
superspace. The results obtained could be realized off-shell
because we used the Legendre transformation technique to reduce
one of the space-like coordinates. In so doing, the Lagrangian
acquired auxiliary fields that make the algebra to close
off-shell. Other important result of this technique is the
appearance of central charge transformation that is consequence of
translation with respect to the coordinates reduced with this
technique. It is interesting because in the real physical
situation the fields may carry a dependence on the extra space
dimensions while the action does not need to show such a
dependence. In a forthcoming work, we shall be discussing more
deeply issues related to the central charges and their relation to
topological configurations that may be found in the $N=2$ and
$N=4$ extensions of the Lorentz-violating model.


\end{document}